\documentclass{aa}
\usepackage{graphicx}

\begin{document}

\title{Enrichment of the ICM of galaxy clusters due to ram-pressure stripping}

\author{W. Domainko$^1$, M. Mair$^1$, W. Kapferer$^1$, E. van Kampen$^1$, 
T. Kronberger$^{2,1}$, S. Schindler$^1$, S. Kimeswenger$^1$, M. Ruffert$^3$
and O. E. Mangete$^3$}

\institute{${^1}$Institut f\"ur Astro- und Teilchenphysik, Leopold-Franzens 
Universit\"at
Innsbruck, Technikerstra\ss e 25, A-6020 Innsbruck, Austria\\
http://astro.uibk.ac.at/astroneu/hydroskiteam/index.htm\\
${^2}$Institut f\"ur Astrophysik,
    Universit\"at G\"ottingen,
    Friedrich-Hund-Platz 1,
    D-37077 G\"ottingen, Germany\\
${^3}$ School of Mathematics, University of
Edinburgh, Edinburgh EH9 3JZ, Scotland, UK}

\offprints{\email{wilfried.domainko@uibk.ac.at}}

\authorrunning{Domainko et al.}
\titlerunning{Enrichment of the ICM by ram-pressure stripping}
\date{Received / Accepted}

\abstract{We investigate the impact of galactic mass loss 
triggered by ram-pressure stripping of cluster galaxies on the evolution of the
intra-cluster medium (ICM). 
We use combined N-body and hydrodynamic simulations together with a 
phenomenological galaxy formation model and a prescription
of the effect of ram-pressure stripping on the galaxies. We analyze the effect of
galaxy -- ICM interaction for different model clusters with different
masses and different merger histories.
Our simulations show that ram-pressure stripping can account for 
$\sim$ 10\% of the overall observed level of enrichment in the ICM 
within a radius of 1.3 Mpc.
The efficiency of metal ejection of cluster galaxies depends at the first few Gyr of the 
simulation mainly on the cluster mass and is significantly increased
during major merger events. 
Additionally we show that ram-pressure 
stripping is most efficient in
the center of the galaxy cluster and the level of enrichment drops quite
fast at larger radii. We present emission
weighted metallicity maps of the ICM which can be compared with X-ray
observations. The resulting distribution 
of metals in the
ICM shows a complex pattern with stripes and plumes of metal rich material. 
The metallicity maps can be used to trace the
present and past interactions between the ICM and cluster galaxies.

}

\maketitle


\section{Introduction}

The presence of an embedding medium has an influence on the evolution of 
galaxies. This effect will be most important in high density environments
like those of galaxy clusters. One mechanism how galaxies can lose parts of 
their
inter-stellar medium (ISM) due to interaction with their surroundings is
ram-pressure stripping (Gunn \& Gott \cite{gunn72}). Observations show that
this mechanism is at work in several galaxies in the Virgo cluster 
(e.g. Cayatte et al. \cite{cayatte90}; 
Kenney et al. \cite{kenney04}; Vollmer et al. \cite{vollmer04a}) 
and in the Coma cluster (Bravo-Alfaro et al. \cite{bravo-alvaro00}, 
\cite{bravo-alvaro01}). Galaxies subject to ram-pressure stripping will
lose parts of their ISM which has been enriched with heavy elements by their 
stars.
This will have an impact on the chemical abundance of the surrounding 
intra-cluster medium (ICM). X-ray observations here
revealed that the ICM is highly enriched with heavy elements (about 0.3 solar
in iron: e.g. Fukazawa et al. \cite{fukazawa98}).
With the recent X-ray satellites \textit{Chandra} and \textit{XMM} it is even
possible to study the 2D distribution of the heavy elements (e.g. Schmidt et al. 
\cite{schmidt02}; Furusho et al. \cite{furusho03}; Sanders et al. \cite{sanders04};
Fukazawa et al. \cite{fukazawa04}; Hayakawa et al. \cite{hayakawa04}). 
As heavy elements cannot be produced in the ICM itself, part of this medium
seems to originate from cluster galaxies. Although 
ram-pressure stripping of cluster galaxies will contribute to the chemical
enrichment of the ICM several other mechanisms like galactic winds
(de Young \cite{deyoung78}), galaxy-galaxy interaction (e.g. Kapferer et al
\cite{kapferer05}) or intra-cluster supernovae (Domainko et al. 
\cite{domainko04}) will also play an important role. 
Each mechanism will result in a characteristic distribution of metal rich 
material
in the ICM,  so with the distribution of heavy elements the origin of the 
metals can be constrained. Observational evidence for some 
chemical enrichment due to ram-pressure stripping has been found: high 
metallicity 
regions in the ICM of the galaxy cluster Abell 1060 stripped from cluster 
galaxies were measured (Hayakawa et al. \cite{hayakawa04}). 
Indeed it has been argued that ram-pressure stripping can act as an 
enrichment agent over a long period of cluster evolution (Finoguenov et al. \cite{finoguenov00})
and at low redshifts (Oosterloo \& van Gorkom \cite{oosterloo05}).
Hence the metals
are a good tracer for the present and past interaction processes between
galaxies and the ICM. 

The high level of enrichment in the ICM has triggered a lot of discussion about
the origin of these metals. Several groups here investigated the chemical evolution 
of the ICM with extensive numerical simulations. De Lucia et al. 
(\cite{delucia04}) and Nagashima et al. (\cite{nagashima04}) investigated this 
problem using N-body simulations combined with semi-analytic galaxy formation 
and evolution. They conclude that most of the metals are ejected by supernova
driven galactic winds mainly from the most massive galaxies and they find  
that there is a mild chemical evolution since z = 1. They do not predict the 
distribution
of the metals in the ICM. Tornatore et al. (\cite{tonatore04}) use a smooth 
particle hydrodynamic simulation and emphasize the contribution of 
different supernova types to the enrichment but do not investigate in detail 
the influence of different ejection mechanisms. Their results
indicate that the simulations still lack a feedback mechanism which quenches
star formation and transports metals at low redshifts. 

In contrast to these papers our calculations emphasize the 
transport mechanisms of the enriched
material and we use a shock capturing method for the hydrodynamic treatment
(see also Schindler et al. \cite{schindler05}, Domainko et al. \cite{domainko05}). 
Furthermore we are able to
trace the distribution of the enriched material lost by the cluster
galaxies.  In this paper we investigate the 
impact of the enrichment of the ICM by ram-pressure stripping of cluster 
galaxies. Other processes are studied elsewhere (e.g. Kapferer et al. 
\cite{kapferer05b}, \cite{kapferer05c}). 

The effect of ram-pressure stripping on individual galaxies was studied 
analytically and numerically by
several groups. A first analytical criterion for ram-pressure stripping
was already given by Gunn \& Gott (\cite{gunn72}). According to this criterion
galaxies lose material if the force due to ram-pressure stripping exceeds the
restoring gravitational force of the galaxy. One result of this consideration is
the presence of a stripping radius in ram-pressure affected galaxies outside
which the galactic gravitational potential is to shallow to prevent the ISM from
being stripped. Various authors studied this stripping radius with respect
to acting ram-pressure and galaxy properties (e.g. Abadi et al. \cite{abadi99},
Vollmer et al. \cite{vollmer01}, Roediger \& Hensler \cite{roediger05}). In
general these authors found a reasonable agreement between the analytically
estimated stripping radius and the stripping radius determined by numerical
simulations.
Additionally Roediger \& Hensler (\cite{roediger05}) discussed the most
appropriate analytical description with respect to the numerical results. These
authors supplementary adapt the estimate used by Mori \& Burkert (\cite{mori00})
which compares the thermal ISM pressure with ram pressure. For the case of a
constant ram pressure acting on the galaxy Roediger \& Hensler
(\cite{roediger05}) found the closest coincidence between analytical description
and numerical simulation for the estimate introduced by Mori \& Burkert
(\cite{mori00}). For the case of a time dependent ram pressure like it is the
case for galaxies moving on non-circular orbits through a galaxy cluster,
Vollmer et al. (\cite{vollmer01}) concluded that the Gunn \& Gott
(\cite{gunn72})
criterion adopted to the galactic inclination angle can explain the HI
deficiency of the galaxies in the Virgo cluster. Further numerical simulations
for ram-pressure stripping on galaxy scales were presented 
in several papers 
(e.g. Abadi et al. \cite{abadi99}; Quilis et al. 
\cite{quilis00}; Mori \& Burkert \cite{mori00}, Vollmer et al. \cite{vollmer01},
Toniazzo \& Schindler \cite{toniazzo01} and Roediger \&
Hensler \cite{roediger05}). In contrast to those papers we present 
simulations showing the impact of ram-pressure stripping 
of all cluster galaxies on the
ICM. We obtain simulations on galaxy cluster scale and compute
the efficiency and time dependence of galactic mass loss triggered by 
ram-pressure stripping. In this paper we focus on the effect of 
ram-pressure stripping on the metal enrichment of the ICM. The effect on
the affected galaxies will be shown in van Kampen et al. (in prep.).

This paper is organized as follows: In Section \ref{num} we present the 
numerical method used. In Section \ref{sim} we describe the general simulation
setup and the properties of the model clusters. In Section \ref{res} we show
our results. And in Section \ref{disc} we summarize and discuss the 
results obtained.


\section{Numerical Method}\label{num}

We use the same general simulation setup as described in Schindler et al. 
(\cite{schindler05}).

\subsection{N-body simulations}\label{darkm}

Our starting point is a model for the dark matter distribution within and
around the cluster, for which we use N-body simulations. 
The N-body code is that of Barnes \& Hut (\cite{barnes86}), adapted to include a
halo formation recipe that prevents over merging on small scales
(van Kampen et al. \cite{vankampen99}). This paper assumes 
a $\Lambda$CDM
cosmology, with $\Omega_\Lambda=0.7$, $\Omega_{\rm m}=0.3$,
$h=0.7$, and $\sigma_8=0.93$.
Initial conditions are produced using the constrained random field method 
of Hoffman \& Ribak (\cite{hoffman91}), as implemented by van de Weygaert \& Bertschinger
(\cite{vandeweygaert96}). Constraints are put on the smoothed density field in the
form of peak heights for a Gaussian window of 6 Mpc, which produce
a rich cluster in the center of the box for peaks above three times the r.m.s.
density fluctuation at that scale.
The N-body simulations are obtained within a box
of 46 Mpc (for H$_0 = 70$ km s$^{-1}$ Mpc$^{-1}$) with 64$^3$ particles.
The softening length of the N-body simulation is 20 kpc which is also the
resolution of the dark matter distribution. 
The whole cluster is described by tens of thousands of dark matter
particles, which can easily resolve the radial structure of the cluster as well as substructures.
The halos of the galaxies are represented by groups of particles with a minimum of 10 particles
with a half-mass radius of the smallest halo of 30 kpc.
For these galactic halos we assume an isothermal profile. The galactic dark matter halo 
gravitation is treated without any spatial details by simply applying the gravitational potential.

\subsection{Hydrodynamic treatment}\label{hydrodyn}

For the hydrodynamic part of the simulation we use a shock capturing, grid
based scheme (PPM, Collela \& Woodward \cite{collela84}, Fryxell et al. 
\cite{fryxell89}). 
PPM is a higher order version of the Riemann solver based Godunov method
(Godunov \cite{godunov59}).
This allows us to
study all the dynamical effects in the ICM connected to subcluster mergers
with high accuracy. To achieve high resolution at the cluster center 
where most 
of the stripping is expected to happen and also to trace infalling gasrich
galaxies from outside the cluster, we compute our simulation on multiple 
refined grids (Ruffert
\cite{ruffert92}). 
Radiative cooling of the ICM is treated according to the cooling 
functions given by Sutherland and Dopita (\cite{sutherland93}). 
Metallicity is used as a tracer to follow the gas which
is lost by the cluster galaxies. Metallicity is handled as a single
quantity therefore we do not distinguish between different chemical 
elements (like iron and $\alpha$ elements) but assume solar composition.
Additionally to the bulk motion of the ICM, diffusion should also further disperse the 
stripped gas. In the presented paper we are mainly interested in the distribution of the
enriched material therefore only the effect of diffusion on heavy nuclei is investigated. 
Diffusion strongly depends on the mass and the charge of the involved nuclei (see Lang
\cite{lang99}, Chapter 3.1.2 and 3.1.5). The most
prominent detection of metals in the ICM is connected to the presence of iron. 
Diffusion will spread iron in the ICM within the Hubble time on length scales of less than 100
kpc. Therefore this process is inefficient in further distributing iron on galaxy cluster 
scales.

\subsection{Galaxy formation}\label{galform}

The halo formation recipe that is part of the N-body code allows us to 
easily generate halo merging histories, and also the merging histories of the galaxies
occupying the halos. Simple prescriptions for gas cooling, star formation and feedback
mechanisms make up our phenomenological galaxy formation model, which is an improved
version of the model of van Kampen et al. (\cite{vankampen99}). 
Important for the context of this paper is the inclusion of a disk model, which
is similar in nature to that of Mo, Mao \& White (\cite{mo98}), but with the inclusion of
an observed threshold for star formation in disks (Kennicut \cite{kennicutt89}; 
Martin \& Kennicutt \cite{martin01}).
The adoption of the Mo, Mao \& White approximations produce a gas and stellar
disk with the same exponential scale-length, but different cut-off radii and central
densities. Our modeling of the metallicity evolution is described in two papers: in
Kapferer et al. (\cite{kapferer05b}) the coupled circuit of the evolution of the stellar, cold
and hot gas components is given, whereas the chemical evolution of the stellar populations
can be found in van Kampen et al. (\cite{vankampen99}).
The star formation history of each galaxy is constrained by its merger history.
Halo-halo mergers and galaxy-galaxy mergers trigger short bursts of star formation.
Metals which are produced in disk stars can either be added to the cold gas disk  or can be
ejected directly into the hot gas, or into the ICM
for cluster galaxies that have their hot gas halos stripped.
We choose the extreme scenario to add metals produced in disk stars to the
cold gas in the disk which can then be stripped off through ram-pressure stripping. 
We do not replenish the stellar disk from stellar ejecta along the path of a galaxy:
star formation is truncated 'forever' if the whole gas disk is stripped, and the galaxy
ends up as an S0 in case the stellar disk is prominent enough, and as an elliptical if
it is not. In reality, some stellar mass loss will occur also after the 
star formation has ended. Intermediate mass stars will eject material in form of AGB winds 
and supernova Ia ejecta. 

\subsection{Ram-pressure stripping}\label{stripping}

To calculate the mass loss of cluster galaxies due to ram-pressure stripping
we follow the classical Gunn \& Gott (\cite{gunn72}) criterion 
because Vollmer et al. (\cite{vollmer01}) concluded that for non
circular orbits the Gun \& Gott (\cite{gunn72}) criterion additionally extended 
for the inclination angle of the galaxy can explain the HI deficiency of galaxies
located in the Virgo cluster. According to
this criterion material is lost by a galaxy if the force due to ram-pressure, 
exceeds the restoring gravitational force. 
The Gunn \& Gott (\cite{gunn72}) criterion is given by:

\begin{equation}
P_\mathrm{ram} \geq 2\pi \mathrm{G}\sigma_\mathrm{star}\sigma_\mathrm{gas}
\end{equation}

with

\begin{equation}
P_\mathrm{ram}=\rho_\mathrm{ICM}v_\mathrm{gal}^{2}
\end{equation}

$P_\mathrm{ram}$ is the ram-pressure, G is the gravitational constant,
$\sigma_\mathrm{star}$ is the stellar surface 
density, $\sigma_\mathrm{gas}$ is the surface density of the gas, $\rho_\mathrm{ICM}$
is the density of the ICM and $v_\mathrm{gal}$ is the velocity of the galaxy
relative to the ICM. We calculate the mass loss of the cold ISM disks of  
spiral galaxies.
We assume that the galaxies consist of an exponential stellar disk and an 
exponential gas disk. Due to the phenomenological galaxy formation model (see Sect.
\ref{galform}) the scale length of the stellar disk and gas disk are equal but the central 
densities are different:

\begin{equation}
\sigma_\mathrm{star,gas} = \frac{M_\mathrm{star,gas}}{2 \pi R^{2}_{0}}e^{-r/R_0}
\end{equation}


$M_\mathrm{star}$ is the mass of the stars, $M_\mathrm{gas}$ is the mass of the gas and
$R_0$ is the disk scale length.
With these assumptions we can analytically determine a stripping radius 
$R_\mathrm{strip}$:

\begin{equation}
x = 0.5 \times \ln  \left( \frac{G M_\mathrm{star} M_\mathrm{gas} }
{v^{2}_\mathrm{gal} \rho_\mathrm{ICM} 2 \pi R^{4}_{0}}
 \right)
\end{equation}

with

\begin{equation}
x = \frac{R_\mathrm{strip}}{R_0}
\end{equation}

Material which is located outside this stripping radius is then lost by the 
galaxy. 
The validity of this treatment was shown with numerical simulations
and with comparisons with Virgo cluster galaxies by Abadi et al. 
(\cite{abadi99}).
The amount of cold gas which can be kept by the galaxy ($M(R_\mathrm{strip})$)
is:

\begin{equation}
M(R_\mathrm{strip}) = M_\mathrm{gas} 
\left[ 1- \left( x+1
\right)  
\mathrm{e}^{-x} \right]
\end{equation}

Consequently the amount of gas $M_\mathrm{strip}$, which is stripped off the 
galaxy is then:

\begin{equation}\label{mgunn}
\frac{M_\mathrm{strip}}{M_\mathrm{gas}} = \left( x+1 \right) \mathrm{e}^{-x}
\end{equation}

Mass loss due to ram-pressure stripping should in principle also be influenced by the
inclination angle of the affected galaxy. The impact of the inclination angle of the galaxy 
on the stripping process was
investigated by Quilis et al. (\cite{quilis00}) with numerical simulation. Those authors find 
that there is only little dependence of the effectiveness of ram-pressure stripping for galaxies 
which move close to face on
through the ICM. For example a galaxy moving at an inclination of 20$^{\circ}$ to the direction of
motion suffers as much stripping as face on encounters (Quilis et al. \cite{quilis00}).
There are, however, also arguments which suggest a certain impact of the inclination angle on the
stripping process.
The column density of the gas that has to be pushed by the momentum transfer
from the ICM grows with the cosine of the inclination angle. 
Also the area of ram pressure which is 
acting on the unit area of an inclined galaxy scales with the same factor.
In this work, due to the lack 
of more elaborate considerations and for simplicity,
mass loss of galaxies moving
inclined as opposed to face on through the ICM is scaled with 
the cosine of the inclination
angle with respect to the mass loss derived for an uninclined galaxy.
Stripped material is added locally to the ICM and is removed from the galaxies.
Note that this treatment can be regarded as an upper limit of the galactic mass 
loss due to
ram-pressure stripping as it was shown that a minor fraction of stripped 
material might be reaccreated back by the stripped galaxy (see Vollmer et al. 
\cite{vollmer01}).
The stripping condition is evaluated at every time step of the cluster scale 
hydrodynamic simulation. The mass loss of the galaxies is calculated according to the
actual properties of the galaxies as it is derived by the galaxy evolution model (see Sect. 
\ref{galform}). 
We also assume that the gas disk remains truncated after 
a stripping event. This means that galaxies are only further stripped when they
experience increasing ram-pressure.
The stripped ISM in the outer part of the disks should in principle be
replaced by supernovae and stellar winds. However observations show that star 
formation is
truncated in ram-pressure affected cluster galaxies (e.g. Koopmann \& Kenney
\cite{koopmann04}). Hence only a passively aging stellar population with a
significantly reduced mass loss is
present in regions of the galactic disk where the ISM is stripped off.
Although as already mentioned in Sec. \ref{galform}, intermediate mass stars will 
eject some enriched material, we do not take this additional effect into account.

After the stripping radius is reached the mass loss continues due to
Kelvin - Helmholz instabilities (Mori \& Burkert \cite{mori00}). Since Roediger
\& Hensler (\cite{roediger05}) argued that the contribution to the mass loss 
caused by Kelvin - Helmholz instabilities is small, we do not account for
this  mass loss in our further calculations.

Observations of nearby spiral galaxies show that abundance 
gradients are present in their disks (see e.g. Henry \& Worthey
\cite{henry99}). The slopes of the gradients depend on the morphology of the
galaxy (Henry \& Worthey \cite{henry99}). This feature exhibits a 
certain influence on the enrichment of
a surrounding medium due to ram-pressure stripping. 
For the case of galactic metallicity gradients which flatten in the outer parts of the
galaxy we expect only a limited effect on the proposed enrichment process. In this case the mean
abundance of the gas is close to the abundance of the bulk of the gas which is
located at large galactic radii. But this issue certainly demands further analysis.
As a first approach we do not consider galactic abundance gradients in our
simulations. In future more refined studies we will investigate the actual
impact of galactic abundance gradients on the ICM enrichment.

\subsection{Supersonic galaxies}

In our simulations several galaxies move with a velocity relative to the 
embedding medium which is higher than the local sound velocity. In this
case of supersonic galaxies the classical Gunn \& Gott (\cite{gunn72})
approach has to be modified. Supersonic galaxies will form bow shocks in front
of them and the stripping criterion has to be applied to the conditions behind
these bow shocks. Hydrodynamic quantities behind the bow shocks can be 
calculated as functions
of the quantities in front of the bow shock according to the Rankine-Hugoniot
conditions for all Mach numbers $M > 1$.
Using the equation of continuity at the bow shock we can calculate
the ratio between densities and velocities in front of and behind the bow 
shock:

\begin{equation}
\frac{\rho_1}{\rho_2} = \frac{v_2}{v_1}
\end{equation}
 
Here $\rho$ are the densities, $v$ are the velocities and subscript 1 stands
for quantities in front of the shock and subscript 2 stands for quantities
behind the shock. For the ram-pressure in front of and behind the shock this
leads to:

\begin{equation}
P_\mathrm{ram,1,2} = \rho_{1,2}v_{1,2}^{2}
\end{equation}

As the velocities behind the shock are smaller than the velocities in front 
of the
shock also the ram-pressure will be smaller behind the shock due to its 
dependence on the square of the velocity. This means that for supersonic
galaxies ram-pressure stripping is less efficient than the classical 
Gunn \& Gott (\cite{gunn72}) approach would suggest. 
We calculate the Rankine-Hugoniot condition for the actual Mach number
of each individual galaxy. Therefore this treatment is also correct for
supersonic galaxy motion in galaxy clusters with low Mach numbers.
The effect is most
pronounced for very fast galaxies. In the case of $v_\mathrm{gal} \gg c_\mathrm{s}$ with 
$v_\mathrm{gal}$ being the velocity of the galaxy relative to the ICM and $c_\mathrm{s}$ the
local sound velocity and treating the ICM as an ideal gas with $\gamma=5/3$ with $\gamma$ being
the ratio of specific heats, the ram pressure is by a factor of 4 lower
behind the galactic bow shock than in front of the bow shock. 
An additional effect in supersonic ram-pressure stripping could come from
the fact that the bow shock is bent according to the Mach number.
Since our simulations are performed on galaxy cluster scale we
cannot resolve individual galaxies and the form of their bow shocks. 
But it is expected that for low Mach numbers the
opening angle of the bow shock is quite large. 
Therefore we do not take into account the effect of
the bending of the bow shock. It is also important to mention that
Roediger \& Hensler (\cite{roediger05}) found a better agreement between
analytic estimates and numerical simulations of ram-pressure stripping if a
Rankine-Hugoniot correction is applied.


\section{Simulation set up and model clusters}\label{sim}

We run the hydrodynamic simulation on four levels of
nested grids centered on the cluster centers, the largest grid being 
(20 Mpc)$^3$ and the smallest (2.5 Mpc)$^3$.
Each of the nested grids has a resolution of $128^3$ grid cells.
We start the hydrodynamic simulation at a redshift of z = 1 
for numerical reasons and follow the 
evolution of the ICM to a redshift of z = 0. This is the time
interval of interest as ram-pressure stripping is expected to be most
efficient after the galaxy clusters have formed.

In order to test the impact of ram-pressure stripping on different galaxy
clusters we choose model clusters which span a wide range of masses and
evolutionary histories. Our model clusters contain no cD galaxies and do not
feature cooling cores at their centers. The model clusters are as follows: 

\begin{description}

\item[\textbf{Model cluster 1:}] This is the most massive cluster of our 
sample. The final total cluster 
mass at $z=0$ inside a radius of 3 Mpc is $1.3 \times 10^{15} 
\mathrm{M_{\odot}}$. This cluster experiences only small merger events after z = 1.

\item[\textbf{Model cluster 2:}] This is the least massive cluster 
in our sample. The final total cluster 
mass at $z=0$ inside a radius of 3 Mpc is $7.4 \times 10^{14}
\mathrm{M_{\odot}}$. This cluster undergoes a major merger with a mass ratio
of 1:3 at a redshift of 0.5. Additionally it features the infall of several smaller 
subsystems.

\item[\textbf{Model cluster 3:}] This cluster has a mass of $8.7 \times 10^{14}
\mathrm{M_{\odot}}$ inside a radius of 3 Mpc at the final redshift of z = 0. 
It undergoes two minor mergers.

\end{description}


\section{Results}\label{res}

\subsection{Total amount of stripped metals}

The first quantity of interest is the total amount of material and metals 
which is ejected into the ICM by ram-pressure stripping. As ram-pressure
stripping is a process dependent on the environment it is expected that the total
amount of material stripped from the affected cluster galaxies will
depend on the properties of the cluster. Galaxy clusters
with higher masses strip galaxies more efficiently than galaxy clusters
with lower masses. This is due to a higher ICM density
and to a higher velocity dispersion in high mass systems. 

In our simulation we derive the total amount of metals stripped from 
the cluster galaxies. By comparing 
the total mass of metals stripped we
test the dependence of galactic metal loss on cluster mass. 
Indeed we find that the total mass loss is
highest in the high mass cluster (model cluster 1) and lower in the low mass
clusters.
We see that in the high mass cluster in our sample (model cluster 1) the total
amount of metals ejected into the ICM is about four times higher than in the 
model clusters with lower mass.

\begin{figure}[h]
\includegraphics[width=7.2cm, angle=-90]{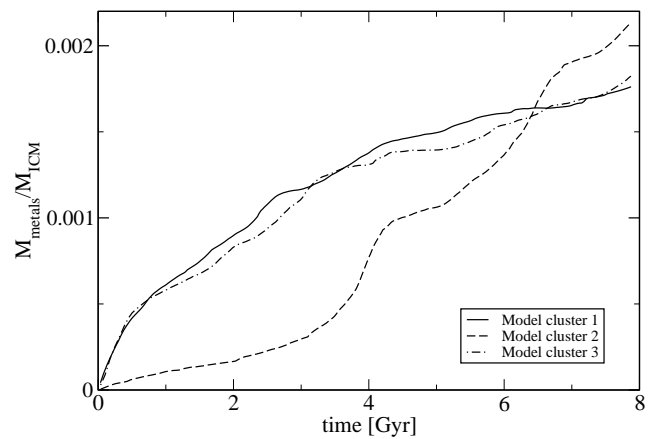}
\caption{Temporal metallicity evolution $M_\mathrm{metals}/M_\mathrm{ICM}$
in the different 
model clusters. The values are given for all galaxies which are located inside
the central (2.5 Mpc)$^3$ of our simulations. For an explanation of the different slopes
for the different model clusters see Sect. \ref{res.loss}.} 
\label{stripm}
\end{figure}

\begin{figure}[h]
\includegraphics[width=7.2cm, angle=-90]{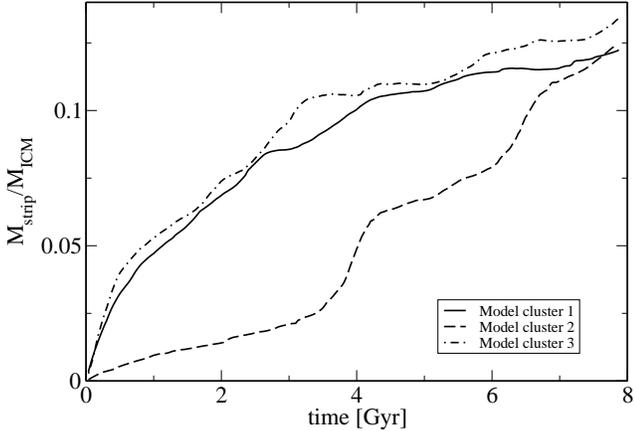}
\caption{Fraction of the ICM which originates from cluster galaxies and is
lost by them due to ram-pressure stripping. The values are given for 
the central (2.5 Mpc)$^3$ of our simulations. For an explanation of the different slopes
for the different model clusters see Sect. \ref{res.loss}.} 
\label{frac}
\end{figure}

Another interesting quantity is the fraction of the ICM which actually
originates from cluster galaxies and is lost by them due to ram-pressure stripping.
To investigate this question, we compare the mass of the hot ICM, which in our 
simulations 
has never been inside any galaxy, to the amount of material which has actually
been lost by the galaxies (see Fig \ref{frac}).

\noindent We find that about 10\% of the ICM in the central (2.5 Mpc)$^3$ of 
our simulation originates from
ram-pressure affected galaxies. 
The fraction of the ICM which originates from cluster galaxies is
comparable for all our model clusters (see Fig. \ref{res.loss}). This result is
expected since in the high mass cluster also the mass of the ICM is larger than
in the clusters with lower mass. Mixing then the ICM in the high mass cluster 
with a larger amount of material 
which is lost by the galaxies results in a comparable fraction of lost ICM 
to primordial ICM as in the low mass clusters.  

\subsection{Mass loss rate due to ram-pressure stripping}\label{res.loss}

\begin{figure}[h]
\includegraphics[width=7.2cm, angle=-90]{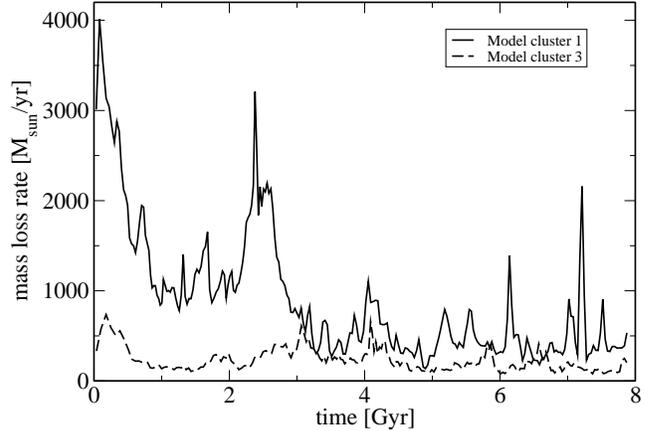}
\caption{Combined mass loss rate of all galaxies as a function of time. 
The values are given 
for the central (2.5 Mpc)$^3$ of our simulations. In this plot the 
influence of cluster mass on the 
combined stripping rate of all cluster galaxies can be seen.
Model cluster 1 is about twice as massive as model cluster 3.} 
\label{rate13}
\end{figure}

\begin{figure}[h]
\includegraphics[width=7.2cm, angle=-90]{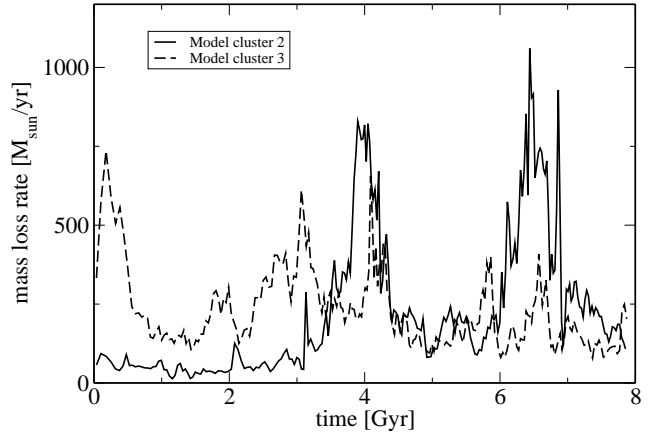}
\caption{Combined mass loss rate of all galaxies as a function of time. 
The values are given 
for the central (2.5 Mpc)$^3$ of our simulations. In this plot the influence of 
the merger histories of the galaxy clusters on the combined stripping rate
can be seen. Model cluster 2 undergoes a major merger at $t$ = 4 Gyr.} 
\label{rate23}
\end{figure}

We now investigate the combined mass loss rate of all cluster galaxies as a 
function of time. It is interesting to mention that the time evolution of the
mass and metal ejection is different for different model clusters 
(see Fig. \ref{stripm}
and Fig. \ref{frac}). With the 
analysis of the total mass loss rate of all cluster galaxies as a function 
of time we further
explore this behavior. The mass loss of all cluster galaxies as a function
of time is shown in Fig. \ref{rate13} and Fig \ref{rate23}. 

We find that in the first 3 Gyr the mass loss rate in the high mass 
cluster (model cluster 1) is about an order of magnitude higher than the mass 
loss rate in low mass clusters.
There is also a prominent peak at the start of the simulation in model
cluster 1. This first peak
originates from the mass loss of gas rich galaxies which experience
the effect of ram-pressure for the first time by switching on 
the calculation. 
After 3 Gyr the mass loss of the galaxies in the high mass cluster
drops significantly. We interpret this as an effect of cluster mass.  
Stripping in the high mass cluster can reduce the 
size of the gas disks of galaxies quite fast and then this mechanism 
becomes less
efficient (see Fig. \ref{rate13}). 

The situation is very different for the 
merging cluster (model cluster 2). 
In the case of a cluster experiencing mergers the prehistory of
ram-pressure is also followed within the infalling subsystems.
In this system mass loss starts at a quite small 
rate but between 3 Gyr and 5 Gyr the rate rises by an order of magnitude.
In this time interval also the major cluster merger happens. Here we see the 
influence of the merger history on the evolution of the mass loss rate due to
ram-pressure stripping (see Fig \ref{rate23}).
At about 6 Gyr several galaxy groups fall towards the cluster center and
the mass loss rate increases again.
Subcluster merger can raise the stripping rate by an enhanced velocity 
dispersion, increased ICM density and due to infall of gas rich galaxies onto
the main cluster. The strongly enhanced mass loss rate during subcluster
mergers is also the reason for the quite different slopes for model cluster 1
and 3
and model cluster 2 in Fig. \ref{stripm} and Fig. \ref{frac}.
The influence of cluster mergers on ram-pressure stripping of cluster 
galaxies is further analyzed in Mair et al. (in prep.).
 
Finally we note that model cluster 3
features a similar mass loss rate of its galaxies (with some scatter due to 
small merger events) over the entire simulation
time of 8 Gyr. Hence on cluster scale ram-pressure stripping is a process which
can act over a quite long period of time.
This supports the idea of some enrichment over the whole redshift range of cluster 
evolution
by ram-pressure stripping (Finoguenov et al. \cite{finoguenov00}).

\subsection{Contribution to the chemical enrichment of the ICM}

Gas originating from cluster galaxies is enriched with heavy elements due to
feedback processes from stars and supernovae. If a large amount of this gas is
lost by the galaxies, the ICM into which this gas is mixed will also 
be enriched
up to a certain level. In this section we investigate the contribution of
ram-pressure stripping on the  
chemical evolution of the ICM. Tornatore et al.(\cite{tonatore04}) 
have argued that some galactic mass loss and metal transport 
at low redshifts is needed
to explain the overall chemical abundance of the ICM. Ram-pressure stripping 
is a promising candidate which could account for the necessary mass transfer 
away from the cluster galaxies.

We explore the capability of ram-pressure stripping for enrichment at low 
redshifts. The mean observed 
chemical abundance of the ICM is about 0.3 solar. For comparison with the 
observed values we derive the mean 
chemical abundance in the central (2.5 Mpc)$^3$ of our simulation.
The final level of enrichment 
is 0.03 solar for model cluster 1, 0.05 solar for model cluster 2 and
0.04 solar for model cluster 3. This corresponds to a
contribution of 10 -- 15\%
to the observed overall enrichment of the ICM, depending on cluster mass and 
evolutionary history of the cluster.
In case of the galaxy cluster which experiences a major merger (Model
cluster 2), the enrichment happens later during the cluster evolution in
comparison to the non-merging cluster (e. g. Model cluster 1). Here it is 
interesting to note that the fraction M$_{\mathrm{metals}}$/M$_{\mathrm{ICM}}$  
(see Fig. \ref{stripm}) is rising faster then the fraction
M$_{\mathrm{strip}}$/M$_{\mathrm{ICM}}$ (see Fig. \ref{frac}) for model
cluster 2. This effect results from the internal metallicity evolution of the
cluster galaxies. Material which is stripped later from galaxies is enriched to
a higher level since stellar feedback is acting for a longer period of time on
this gas.

\subsection{Metallicity profiles}

\begin{figure}[h]
\includegraphics[width=8.7cm]{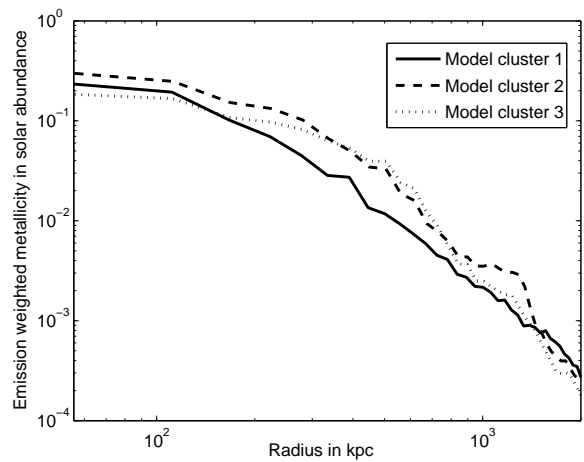}
\caption{Emission weighted metallicity profiles of the three model clusters
at a redshift of z = 0.
Ram-pressure stripping can account for the observed metallicity in the central
100 kpc in all simulations, but at larger radii other processes than ram-pressure
stripping have to contribute to the enrichment of the ICM.} 
\label{profile}
\end{figure}

For a better comparison of the results from our simulations with X-ray observations
we derive emission weighted metallicity profiles. 
These are shown for the three model clusters in Fig. 
\ref{profile}. The profiles for all three model clusters 
are remarkably similar and do not show a strong signature of the different merger 
histories of the different systems. 

In the calculated profiles we clearly see the highest level of chemical 
enrichment of the ICM at the cluster center. At the cluster center the 
environmental impact on the galaxies will be most important so it is expected that
there the enrichment is most pronounced. In the central 100 kpc ram-pressure 
stripping can account 
for an enrichment of about 0.2 solar which is close to the 
observed value. At larger radii the chemical abundance drops very quickly and
other processes than ram-pressure stripping must contribute to the chemical evolution of the 
ICM. 

\subsection{Distribution of enriched material}

\begin{figure*}
\centering
\includegraphics[width=\textwidth]{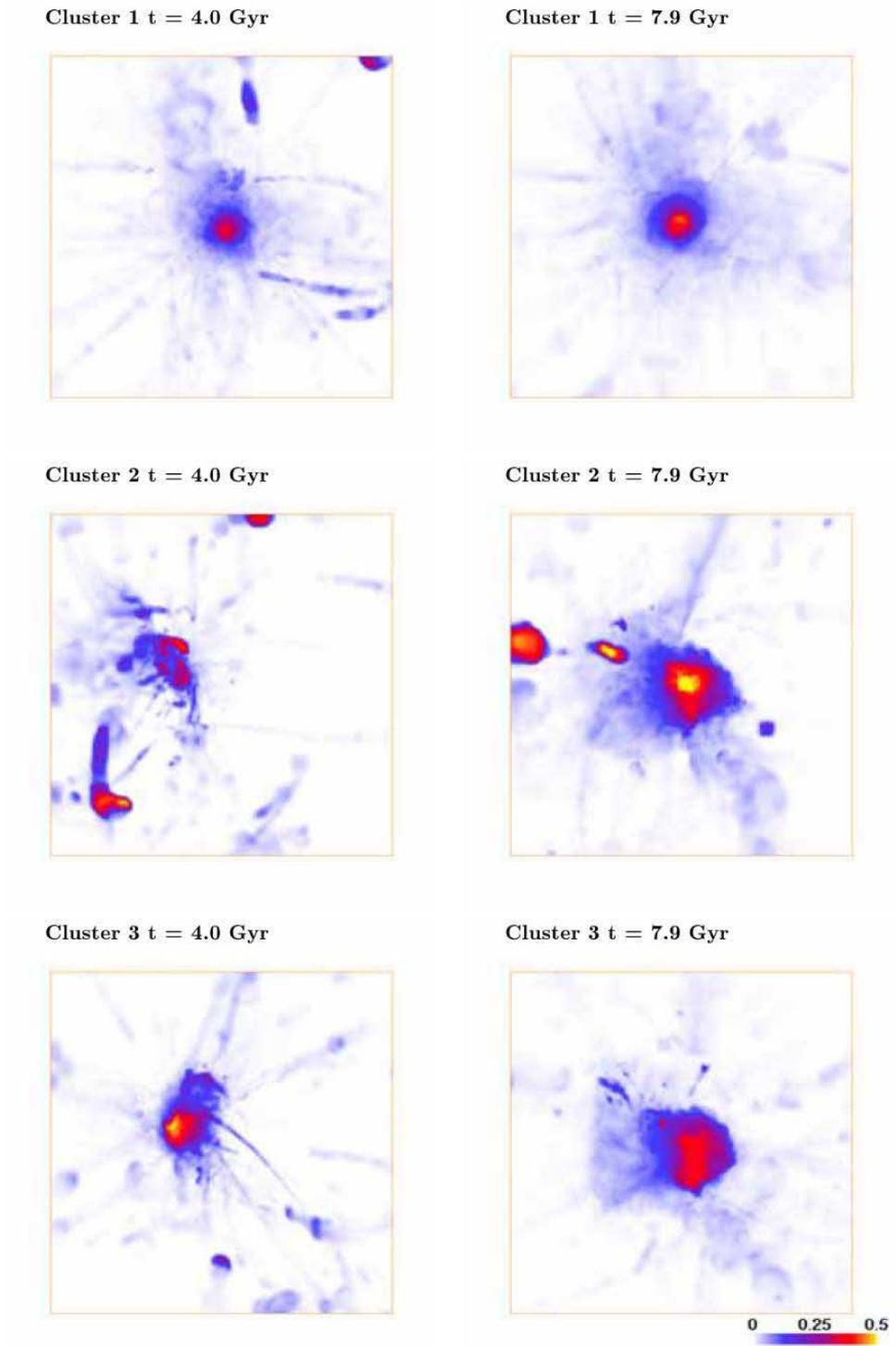}
\caption{Emission weighted metallicity maps of the model cluster. The size of 
the boxes is 5 $\mathrm{Mpc}$ on a side. The level of enrichment is 
given in solar units.  
}
\label{image}
\end{figure*}

Metallicity can be used as a tracer to identify the location of the gas which
was lost by cluster galaxies. To investigate the distribution of the 
enriched material and to compare our simulations with observed galaxy clusters, 
we obtain emission weighted metallicity maps. For a discussion on the effect of diffusion
on the metallicity distribution see Sect. \ref{hydrodyn}. 
In contrast to the full 3D distribution of the metal rich areas, enriched
material located in high density regions is much more pronounced
in emission weighted metallicity maps. 
The emission weighted metallicity
maps clearly show a complex pattern of the chemically enriched material
(see Fig. \ref{image}). 
A nonuniform distribution of the enriched material is in good agreement with
recent observations (e.g. Hayakawa et al. \cite{hayakawa04}).

In all the model clusters we see stripes of metal rich material which were left
behind by infalling galaxies. Some of them are even several 
hundred kpc long. It
is interesting to note the similarity of these stripes with the plume of stripped gas 
observed at
the center of the Virgo cluster in HI (Oosterloo \& van Gorkom 
\cite{oosterloo05}) and with the trails of ionized gas behind two galaxies
in Abell 1367 (Gavazzi et al. \cite{gavazzi01}). In contrast to these observations, 
material in our
computed enriched stripes is already mixed with the hot ICM.
Evaporation of stripped gas clouds at the center of our model cluster 
($T_\mathrm{ICM}$ $\sim$ 10$^8$ K,
$n_\mathrm{ICM}$ $\sim$ 10$^{-3}$ cm$^{-2}$) can occur within 10$^7$ yr (Vollmer et al.
\cite{vollmer01}),  which is the typical time step of the hydrodynamic simulation.
Stripped gas from cluster galaxies is indeed seen in the hot (T = 10$^8$ K)
ICM of the Coma cluster (Finoguenov et al. \cite{finoguenov04}) in X-ray
observations.

The non-merging model cluster 1 shows metal maps with an almost 
spherically symmetric
appearance. There is only little evolution between 4.0 Gyr and 7.9 Gyr
in our simulation (see Fig. \ref{image} two panels at the top).
This is in good agreement with the results of Section \ref{res.loss}. 
In contrast model cluster 2 shows a very complex distribution of the 
enriched material and there is a lot of enrichment happening between a
simulation time of 4.0 Gyr and 7.9 Gyr. The subcluster merger can 
clearly be seen in the panel showing the situation at 4.0 Gyr. The emission
weighted metallicity maps for model cluster 2 are presented in Fig. \ref{image}
in the two central panels. Several stripes of enriched material can
be seen in the maps of model cluster 3 (see Fig. \ref{image} lower two panels).
There is also some evolution up to the end of the simulation at 7.9 Gyr.

Summarizing, our simulations enable us to derive abundance maps of the ICM. With the
help of these maps we can identify the origin of highly enriched regions
which have recently been observed with X-ray observations (e.g. Hayakawa et al.
\cite{hayakawa04}). Furthermore we can investigate the location and the fate 
of the stripped material and compare this to up to date X-ray data
(e.g. Finoguenov et al. \cite{finoguenov04}).


\section{Summary and discussion}\label{disc}

In the present paper we investigate the chemical enrichment of the ICM
due to ram-pressure stripping of cluster galaxies from z = 1 to z = 0.
 We find that ram-pressure 
stripping plays a significant role in this stage of cluster evolution.
In particular we investigate
the efficiency, time dependence and spatial distribution of this process:

\begin{itemize}

\item More than 10\% of the ICM originates from gas of cluster galaxies 
which was lost 
by them by ram-pressure stripping.

\item Enrichment by ram-pressure stripping contributes to the overall 
metal enrichment of the ICM by about 10 -- 15\% within a radius of 1.3 Mpc. 
It is more difficult to enrich high
mass clusters with high ICM densities than low mass clusters with lower
ICM densities. This effect is hardly compensated by the higher efficiency
of ram-pressure stripping in high mass systems.

\item The total mass loss rate of all cluster galaxies depends on the cluster
mass. High mass clusters can reduce the gas disks of their galaxies very quickly
and enrichment of the ICM by ram-pressure stripping can take place within a
few Gyr.

\item Enrichment by ram-pressure stripping persists over the entire simulation 
time and can therefore account for some late enrichment of the ICM. This
effect is more pronounced in low mass systems.

\item Cluster mergers have an important impact on enrichment by ram-pressure
stripping.
Subcluster mergers can raise the total mass loss of all cluster galaxies
significantly

\item Ram-pressure stripping can account for the total observed level of enrichment
within the central 100 kpc but at larger radii other processes have to contribute
to the chemical evolution of the ICM.

\item Metallicity gradients are not very well suited to investigate the complex
distribution of metals in the ICM. Metal maps are much more useful to 
investigate the present and past interactions between the ICM and the cluster 
galaxies.

\item We find an inhomogeneous distribution of the enriched material in the
ICM. This is in agreement with recent observations (e.g. Hayakawa et al.
\cite{hayakawa04})

\item In our simulations we see stripes of metal rich material which were left
behind by ram-pressure affected galaxies. We note the remarkable similarity 
of the shape of
these stripes with the plume of stripped material observed in the Virgo cluster
(see Oosterloo \& van Gorkom \cite{oosterloo05}).   

\end{itemize}

It seems that merger activity of the galaxy cluster has an important impact
on the enrichment of the ICM due to ram-pressure stripping of cluster
galaxies. Studying a larger sample of merging clusters will lead to a better 
understanding of this problem. In the future we plan to investigate the effect
of various merger scenarios on the chemical evolution of the ICM.

In forthcoming more refined studies we will also investigate the impact
of additional effects like abundance gradients in the galaxies, stripping due to
Kelvin - Helmholz instabilities, shapes of bow shocks of supersonic galaxies and
different galaxy evolution scenarios on the enrichment of the ICM due to
ram-pressure stripping.


\begin{acknowledgements}
The authors want to thank an anonymous referee for insightful and substantial comments 
that greatly improved the clarity and content of this paper.
This work was supported by the Austrian Science Foundation FWF under
grant P15868, UniInfrastruktur 2004, Tiroler Wissenschaftsfonds,  
the AUSTRIAN GRID, a graduate scholarship from the University of
Innsbruck, DFG grant Zi 663/6-1, the European Commission through 
grant number HPRI-CT-1999-00026
(the TRACS Program at EPCC) and by the European
Commission's Research Infrastructures activity of the Stucturing the
European Research Area programme, contract number RII3-CT-2003-506079
(HPC-Europa). MM acknowledges the support of 
a Doktoratsstipendium of the LFU Innsbruck. EvK is supported by the FWF
through grant P18416. 
Edmund Bertschinger and Rien van de Weygaert are acknowledged
for providing their constrained random field code, Joshua Barnes and Piet Hut
are acknowledged for allowing use of their treecode.
\end{acknowledgements}

\end{document}